\PassOptionsToPackage{numbers}{natbib}
\documentclass{article}

    \usepackage[preprint]{neurips_2024}

\usepackage[utf8]{inputenc} 
\usepackage[T1]{fontenc}    
\usepackage{hyperref}       
\usepackage{url}            
\usepackage{booktabs}       
\usepackage{amsfonts}       
\usepackage{nicefrac}       
\usepackage{microtype}      
\usepackage{xcolor}         

\usepackage{algorithm}
\usepackage{algpseudocode}
\usepackage{graphicx}
\usepackage{siunitx}

\usepackage{subcaption}
\usepackage{float}
\usepackage{caption}
\usepackage{adjustbox}
\DeclareSIUnit{\molar}{M}

\title{Explainable Deep Learning Framework for SERS Bio-quantification}

\author{
  Jihan K. Zaki\\
  Melville Laboraratory for Polymer Synthesis\\
  Yusuf Hamied Department of Chemistry\\
  University of Cambridge\\
  Lensfield Rd, CB2 1EW \\
  \And
  Jakub Tomasik \\
  Department of Chemical Engineering and Biotechnology \\
  University of Cambridge\\
  Philippa Fawcett Drive, CB3 0AS \\
  \And
  Jade A. McCune \\
  Melville Laboraratory for Polymer Synthesis\\
  Yusuf Hamied Department of Chemistry\\
  University of Cambridge\\
  Lensfield Rd, CB2 1EW \\
  \And
  Sabine Bahn \\
  Department of Chemical Engineering and Biotechnology \\
  University of Cambridge\\
  Philippa Fawcett Drive, CB3 0AS \\
  \And
  Pietro Liò \footnotemark[1] \\
  Department of Computer Science and Technology\\
  University of Cambridge\\
  15 JJ Thomson Ave, CB3 0FD \\
  \texttt{pl219@cam.ac.uk} \\
  \And
  Oren A. Scherman \thanks{Corresponding author}\\
  Melville Laboraratory for Polymer Synthesis\\
  Yusuf Hamied Department of Chemistry\\
  University of Cambridge\\
  Lensfield Rd, CB2 1EW \\
  \texttt{oas23@cam.ac.uk} \\
}

\begin{document}

\maketitle

\begin{abstract}
 Surface-enhanced Raman spectroscopy (SERS) is rapidly gaining attention as a potential fast and inexpensive method for biomarker quantification, which can be combined with deep learning methodology to elucidate complex biomarker-disease relationships. Current standard practices in computational SERS analysis are substantially behind the state-of-the-art machine learning approaches; however, present challenges of SERS analysis could be effectiely addressed with a robust computational framework. Additionally, there is a particular need for improved model explainability for SERS analysis. While current methods are capable of providing global explainability, they are insufficient in assessing the lower level contexts where other factors, such as confounding factors, outliers and measurement errors, or different predictors of the same outcome variable, could affect prediction outcomes. This study presents a novel framework for SERS bio-quantification rooted in a three-step process, including spectral processing, analyte quantification, and model explainability. To this end, a serotonin quantification task in a urine medium was assessed as a model task with 682 SERS spectra measured in a micromolar range using gold nanoparticles and cucurbit[8]uril chemical spacers. A denoising autoencoder was developed for spectral enhancement, convolutional neural networks (CNN), and vision transformers were utilized for biomarker quantification. Lastly, a novel context representative interpretable model explanations (CRIME) method was developed to suit the current needs of SERS mixture analysis explainability. Serotonin quantification was most efficient in denoised spectra analysed using a convolutional neural network with a three-parameter logistic output layer (Validation: mean absolute error (MAE) = \SI{0.24}{\micro\molar}, mean percentage error (MPE) = 15.00\%, Test: MAE = \SI{0.15}{\micro\molar}, MPE = 4.67\%). Subsequently, the CRIME method revealed the CNN model to present six unique prediction contexts, of which three were associated with serotonin. The proposed framework could unlock a novel, untargeted hypothesis generating method of biomarker discovery considering the rapid and inexpensive nature of SERS measurements, and the potential to identify biomarkers from CRIME contexts, which should be validated in a clinical setting.
\end{abstract}

\section{Introduction}

Deep learning methods are increasingly being used in biomarker research, as yet-to-be discovered relationships between biomarkers and disease outcomes increase in complexity with expanding numbers of biomarkers investigated\cite{Mathema2023}. Surface-enhanced Raman spectroscopy (SERS) presents a novel molecular assaying method and allows for the delivery of consistent, accurate, and sizable data that can be utilized with deep learning methods. The technique capitalizes on ‘hot-spots', localized regions of intense optical fields, created by the aggregation of noble metal nanoparticles\cite{Langer2020,Pilot2019}. These nanoparticles offer a robust platform for in situ analysis within liquid media, rendering SERS a practical choice for broad applications. There are a set number of challenges in SERS-based analyte detection, which if solved could unlock the significant potential of the method. These primarily relate to the reproducibility, and readability of spectra, particularly in mixtures\cite{Xiong2014}. Namely, inherent variability in SERS affects the signal intensities of spectra in repeat measurements, and the complexity of biological media measured, alongside potential \textit{intra}- and cross-individual variation in molecular patterns cloud the spectra through biological noise\cite{Zong2018}. While experimental developments are seeing improvements in the applicability of SERS, computational frameworks must be developed to supplement and enable the application of SERS in clinical practice. 

The SERS domain is far behind current state-of-the-art practices developed in the field of machine learning, and there are several promising methodologies yet to be integrated to SERS analyses. SERS analysis has relied on traditional dimensionality reduction methods to reduce the variations of the spectra and to account for the high noise levels of SERS spectra particularly in biological samples. Principal component analysis and discriminant analysis (PCA-DA) and partial least squares regression (PLSR) have been the \textit{de facto} industry standard within the SERS domain\cite{Kasera2014,Lussier2020,Kim2020,Czaplicka2021,Cao2023,Akbar2022,Tahir2024}, however while these methods can be useful for feature extraction, the increased complexity of biological media can hinder predictions. Over the last 5 years, the application of convolutional neural networks (CNN) \cite{LeCun1998} for SERS analysis has become more common\cite{Tang2021,Lussier2019,Huang2021,Thrift2019}. Nevertheless, the application of CNNs has been predominantly applied with established model architectures with limited exploration of domain specific modified layers. Moreover, while transformers have revolutionized various domains of machine learning by enabling models to handle sequential data with long-range dependencies effectively, their application in SERS spectral analysis remains underexplored. To date, we were able to source only two studies where vision transformer-models (ViT) \cite{Dosovitskiy2020} were applied in SERS based applications\cite{Li2023, Tseng2023}. Similarly to transformer models, the application of autoencoders in the field of SERS analysis is scarce and is primarily deployed for improved feature extraction\cite{Ciloglu2021,Ciloglu2022}, despite their strong promise as robust denoising models. Most notably, a majority of SERS studies fail to show adequate model explainability. Without significant exploration of the models selected features, it cannot be determined without reasonable doubt that the predictions are due to the intended signal, or confounders or other sources of sample bias. 

This study aims to develop computational methods to mitigate the described primary challenges of SERS, namely the variability between spectra, and the effects of biological noise on measurements. To this end, the present study proposes a complete and up-to-date SERS analysis framework enabling robust bio-quantification, and explainability. The assessment of urinary biomarkers for mental health disorders was selected as the model system to assess the strengths of SERS. Urine is non-invasive, and easy to obtain in large quantities, and shows significant potential as a biomarker source with over 5000 analytes identified to date\cite{Wishart2022}. In turn, serotonin plays a crucial role in regulating mood, emotion, and sleep, among other physiological functions, and imbalances in serotonin levels have been closely associated with various mental health disorders and has been long hypothesized to be a causal factor in major depressive disorder (MDD), anxiety, and schizophrenia\cite{Lin2014,Coppen1967,Jauhar2023,Gordon2004,Eggers2013}. In brief, this study aims to extend the clinical applicability of SERS in three main directions. First, it seeks to computationally mitigate inherent biological and method-based variations in SERS. Second, it aims to explore deep learning models for more accurate targeted analyte quantification. Lastly, it aims to propose a framework for explaining the decision making process of the developed models, which could identify all contexts in which a model uses different predictors to reach the desired target outcome.

\section{Methods}

\subsection{Dataset Preparation}
\begin{figure}[htbp]
    \centering
    \includegraphics[width=1\linewidth]{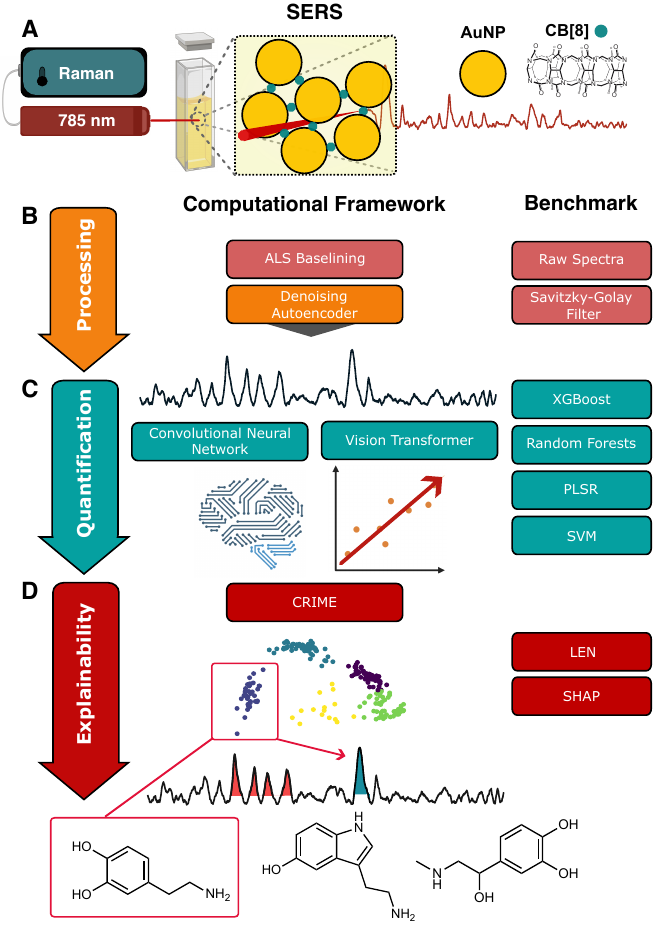}
    \caption{\textbf{SERS deep learning framework development pipeline.} Illustrated are the SERS measurement process applied (A), and the computational framework pipeline. Benchmark comparisons of alternative methodology are presented on the right. Preprocessing methods (B) are marked in orange and light red, quantification methods (C) in blue, and explainability methods (D) in dark red. Asymmetric least squares (ALS) baselining is applied to all spectra prior to assessing the framework or the benchmarks. SERS = surface-enhanced raman spectroscopy, AuNP = gold nanoparticle, CB[8] = cucurbit[8]uril, CNN = convolutional neural network, XGBoost = extreme gradient boosting trees, PLSR = partial least squares regression, SVM = support vector machines, CRIME = context representative interpretable model explanations, LEN = logic explained networks, SHAP = Shapley additive explanations. }
    \label{fig:1}
\end{figure}

The study design is summarized in \textbf{Figure \ref{fig:1}}. The dataset assessed consisted of 318 SERS spectra measured in a lyophilized urine medium and 364 SERS spectra measured in a water medium. Samples were measured using a \SI{785}{\nano\meter} laser and cucurbit[8]uril (CB[8]) spacers (\SI{0.9}{\nano\meter}) with \SI{60}{\nano\meter} gold nanoparticles, as visualised in \textbf{Figure \ref{fig:1}A}. Both urine- and water-medium samples were spiked with three key neurotransmitters: epinephrine, dopamine, and serotonin, with concentrations ranging from 0 to \SI{9}{\micro\molar}. Serotonin was used as the target analyte. This was due to the lyophilized urine medium containing varying endogenous concentrations of epinephrine and dopamine resulting in their unknown absolute concentrations. The SERS spectra were shortened to feature a relevant range of Raman shifts from 300 to 2000 cm\textsuperscript{-1}, resulting in a total of 842 datapoints per spectrum. The specific concentrations of the neurotransmitters in each sample are presented in \textbf{Table \ref{neurotransmitter-table}}. Prior to assessment, spectra were processed using the asymmetric least squares (ALS) algorithm \cite{Paul2004} for baseline correction with pre-specified parameters ($\lambda$=1000, p=0.1, n=10). Following the correction, the data was normalized to intensities between 0 and 1.

\begin{table}[h]
  \caption{\textbf{Added concentrations and number of spectra for all three neurotransmitters in both water and urine backgrounds.} *Samples D, E, and F were not present in the water dataset, and sample M was not present in the urine dataset. Sample U represents a baseline measurement with no added or measured concentrations of the neurotransmitters. EPI = Epinephrine, DA = Dopamine, 5-HT = Serotonin, n = number of spectra.}
  \label{neurotransmitter-table}
  \centering
  \small
  \renewcommand{\arraystretch}{1}
  \setlength{\tabcolsep}{3pt}
  \begin{tabular}{cccccc}
    \toprule
    \textbf{Sample} & \textbf{EPI} & \textbf{DA} & \textbf{5-HT} & \textbf{Water} & \textbf{Urine} \\
    \cmidrule(r){2-6}
                    & (\SI{}{\micro\molar}) & (\SI{}{\micro\molar}) & (\SI{}{\micro\molar}) & (n) & (n) \\
    \midrule
    A & 2 & 0 & 0 & 25 & 22 \\
    B & 0 & 2 & 0 & 17 & 21 \\
    C & 0 & 0 & 2 & 22 & 22 \\
    D & 3 & 0 & 7 & 0* & 26 \\
    E & 0 & 8 & 3 & 0* & 28 \\
    F & 7 & 3 & 0 & 0* & 21 \\
    G & 3 & 2 & 7 & 99 & 26 \\
    H & 1 & 1 & 9 & 38 & 22 \\
    I & 2 & 8 & 3 & 37 & 22 \\
    J & 6 & 9 & 1 & 93 & 23 \\
    K & 7 & 3 & 2 & 11 & 23 \\
    L & 9 & 6 & 4 & 11 & 21 \\
    M & 3 & 3 & 3 & 11 & 0* \\
    U & 0 & 0 & 0 & 58 & 41 \\
    \bottomrule
  \end{tabular}
\end{table}

\subsection{Denoising Autoencoder}
Following baseline correction and normalization, the spectra were denoised using a denoising autoencoder. For conventional autoencoders, an encoder neural network is trained to convolute input data into a latent space, and simultaneously a decoder neural network is trained to restructure the original data from the latent transformation. A denoising autoencoder differs by attempting to reconstruct clean outputs from a latent space formed by encoding noisy data \cite{Vincent2008}, which could prove useful in SERS applications with significant biological noise. A simple denoising autoencoder was trained using full water-medium spectra consisting of 364 spectra with 937 datapoints using an 80:20 train-test split, with the urine background samples incorporated as noise. Noisy data were generated using urine background data (Table \ref{neurotransmitter-table}: sample U), where a randomly selected measurement was overlayed to each clean spectra following baseline correction but before normalization. The denoising autoencoder was implemented in TensorFlow. The encoder comprises two dense layers. The first layer has 2 × 200 units and utilizes a ReLU activation function. The second layer further compresses the data into an encoding space of dimension 200, with ReLU activation. Symmetric to the encoder, the decoder also consists of two dense layers. The first layer expands the encoded data back to 2 × 200 dimensions using ReLU activation. Finally, the second layer reconstructs the data to its original dimension (937) using a sigmoid activation function. The model was compiled using mean squared error (MSE) as the loss function and optimized using the Nesterov-accelerated Adaptive Moment Estimation (Nadam) optimizer. Training was conducted for 128 epochs with a batch size of 32. Both training and testing was performed on noisy data to facilitate the denoising objective. The quality and utility of denoised spectra was subsequently evaluated through effects on performance in quantification models.

\subsection{Quantification models}
The quantification of serotonin was primarily evaluated using state-of-the-art neural network models, as shown in \textbf{Figure \ref{fig:1}C}. The two model types applied to analyse the spectra were the CNN and the ViT, with custom SERS-specific layers evaluated for the CNN. Both the CNNs and the ViT models were implemented in TensorFlow and designed to adapt to SERS spectral data. A core CNN architecture comprised of ReLu and Tanh-ReLu paired 1-D convolution layers. The core architectures of the quantification models are described in more detail in the \textbf{Supplementary Material section \ref{sec:NNQM}.} Three variants of the CNN were evaluated with varying additional custom layers. These included the base CNN with a linear final activation output layer, a CNN with a modified custom Three-Parameter Logistic (3PL) activation function, and a CNN model with inherent scale-adjusting capabilities through scaling layers. 

The scale-adjusting CNN model was developed with two unique scaling layers implemented. These were a multi-scale assessing layer,  and a local scaling layer. Both layers were utilized prior to the half-peak ReLu layer in the core CNN architecture. The multi-scale layer captures features from the input \(X\), with three layers sized 8, 25, and 50, each with 8 filters. Each convolutional operation is defined as \( C_i(X) = W_i \ast X + b_i \), where \(\ast\) denotes a convolutional operation, \(W\) the weight, and \(b\) the bias. To assess the spectra at different scales simultaneously, the output of each convolutional layer is combined following the convolutional operations along the feature dimension. The local peak scaling layer in turn was developed to scale regions of the spectra which were assessed not to be relevant to the outcome variable, identified from the reference spectra of the pure compounds in water. The layer applies a set of scaling factors \(s_j\) unique to the number of pre-registered regions of interest (or non-interest), which are defined by start and end indices \(a_j, b_j\), in the spectra. The scaling operation for each region is expressed as: \(S_j(X) = X_{aj:bj} \odot s_j\), where \( \odot \) denotes the element-wise multiplication. The output of the scaling operation is concatenated to reconstruct the spectra with scaled regions. The modified output layer assessed in both custom layer CNN models utilizes a Three-Parameter Logistic (3PL) activation function. The ViT architecture in turn consisted of an embedding layer with a patch size of 25 matching the CNN architectures initial layer, with a hidden size of 64 and a dropout rate of 0.1. Subsequently, the architecture consisted of 6 transformer blocks with 6 multi-head perceptron’s. The transformer blocks each employed Gaussian Error Linear Unit (GELU) activation functions.

Each model was evaluated in both the raw and denoised data incorporating unseen spectra as well as repeat spectra, with spectra defined as repeat if separate measurements of the specific sample were used in training either the denoising autoencoder or the quantification models. Repeat spectra were split into training and validation sets with a 90:10 split, and furthermore measurements taken from an unseen serotonin free sample (sample F) were included in the validation set. The remaining unseen samples (D, and E) were included exclusively in the test set. Final spectra counts for both datasets consisted of 218 training spectra, with a validation set of 46, and a test set of 54 spectra. Hyperparameter tuning and architecture search for both the CNN variants and the ViT was conducted iteratively, guided by the model's performance on the validation set. Each model variant for both the CNNs and the ViT was trained 100 times, with an ensemble average used for evaluation. Both model types were optimized using the Adaptive Moment Estimation (Adam) algorithm with a learning rate of 0.001, a batch size of 64, and 256 epochs, and compiled with a mean absolute error (MAE) loss function. Additional evaluation metrics included mean squared error (MSE) and mean percentage error (MPE). Early stopping with a patience of 64 was employed to mitigate overfitting, and model checkpoints were saved for epochs that minimized validation loss. Reproducibility was ensured by setting random seeds for TensorFlow, NumPy, and the train-test split. Of the 100 trained models in the ensemble, the model with the lowest MAE in the validation set was selected as the final model, which was assessed in the holdout test set. 

\subsection{Context representative interpretable model explanations}
The reliability and explainability of the final quantification model were assessed using the Context Representative Interpretable Model Explanations (CRIME) framework, developed in this study for machine learning interpretations of data with expected contextual prediction clusters. The CRIME framework expands on the widely applied local interpretable model agnostic explanations (LIME) framework\cite{Ribeiro2016}, by assessing model explanations through contexts. Contexts can be defined within this framework as prominent and consistent explanation outcomes across a number of prediction instances. While contexts can have numerous explanations as to why they are prominent from other explanation contexts, they can be broadly categorized to stem from differing sources of prediction reasoning, such as confounding factors, outliers and measurement errors, or different predictors of the same outcome variable. The framework is summarized in \textbf{Algorithm \ref{CRIME}}.

\begin{algorithm}
\caption{Context Representative Interpretable Model Explanations}
\label{CRIME}
\begin{algorithmic}[1]
\State \textbf{Input:} Dataset $X$
\State \textbf{Initialize:} Explained model $\mathcal{M}$
\State \textbf{Initialize:} LIME model $\mathcal{L}$
\State \textbf{Output:} CRIME context explanations $C$
\State Extract local explanations $\epsilon$ using LIME model $\mathcal{L}$ on model $M$ predictions on dataset $X$
\State Initialize CRIME variational autoencoder model $V$ using explanations $\epsilon$
\State Apply CRIME encoder $e$ to explanations $\epsilon$ to assess the initialized latent space $L$
\State Apply K-means clustering algorithm $\mathcal{K}$ to initialized latent space $L$ with $n$ visually plausible contexts
\State Assign context labels $c$ to latent space $L$ points based on clustering
\State Initialize CRIME context explanation array \( C \) of size \( n \)
\For{\( c = 1 \) to \( n \)}
    \State Compute mean LIME explanation \( \varepsilon_c \) from explanation instances \( \epsilon_c \)
    \State \( C[c] \) \(\leftarrow \varepsilon_c \)  
\EndFor
\State \textbf{return} $C$
\end{algorithmic}
\end{algorithm}

The CRIME framework attempts to identify all prediction contexts of the input data space through the latent space of a variational autoencoder (VAE) trained on the LIME predictions of all instances in the available data. The LIME predictions are flattened with regards to perturbation limits and weights prior to input and are subsequently projected to the two-dimensional latent space. The VAE architecture used in this study consisted of a simple encoder, sampler, and decoder. Notably, the architecture can be fine-tuned depending on the individual requirements for the CRIME framework in future applications. Details regarding the VAE of the CRIME method are described in the \textbf{Supplementary Material section \ref{sec:CRIME}}. Following training of the VAE, the latent space is utilized to identify context clusters representing all the possible ways in which the quantification model interprets the input data. The latent space instances are clustered into the final contexts using K-means clustering, and the latent space is visually inspected for selecting the number of clusters. Finally, a mean LIME explanation is assessed through averaging all instances in each cluster to represent the contexts. To identify the defining features of each context representation, normalized LIME feature weights are combined with mean feature values representing the spectral intensities within the context clusters. They are then set in a three-dimensional space, together with normalized feature positions, which are then further clustered into 15 clusters using K-means clustering. Following the clustering of mean spectral feature values, position z-scores, and LIME weights, the clusters are ordered according to the product of their LIME weights and spectral intensities. The five clusters with the highest score are selected to represent the regions of the spectra which contribute most to the contextual predictions. 

Following the identification of the most relevant context prediction regions, the highlighted regions of the mean context spectra are assessed against measured clean spectra of the neurotransmitters known to be present in the mixture. To emphasize the explanation weights in the spectra, both the reference clean spectra, and the mean context spectra are scaled according to the explanation weights in the specific feature location. To determine the cause or identity of the recognized context clusters, the final mean context indicators are compared to the weighted reference spectra using cosine similarity \(S_c\), which is defined in \textbf{Equation \ref{equation2}}.

\begin{equation}
\label{equation2}
S_{\text{cos}} = \frac{\mathbf{A} \cdot \mathbf{B}}{\|\mathbf{A}\| \|\mathbf{B}\|}
\end{equation}

\subsection{Benchmarking}

Each segment of the framework presented in this study is carefully benchmarked against alternative models or methods, previously established benchmarks, or common practices in the SERS domain. 

The utility of the denoising autoencoder was assessed by measuring performance in the raw and denoised data, and additionally, comparing it with fifth order polynomial second Savitzky-Golay derivative processing with a window length of 33\cite{Kasera2014}, which was previously assessed as a reference standard. The developed CNN architecture was benchmarked against simpler architectures without Tanh-ReLu pairing layers. Furthermore, the best-performing quantification model was assessed against methods used previously to quantify neurotransmitter concentrations, as well as other machine learning methods, including partial least squares regression (PLSR), random forests, support vector machines (SVM), and extreme gradient boosting (XGBoost). The primary benchmark for serotonin quantification accuracy is the previous study by Kasera \textit{et al.} 2014. Hyperparameter tuning was determined using grid search and 3-fold cross validation. The searched parameters are included in the \textbf{Supplementary Material section \ref{sec:Benchmarks}}. Final model robustness was tested using perturbation testing. Gaussian noise was added in two ways across the input spectra, first through applying across the entire spectra to assess overall noise tolerance, and second through applying noise to identified relevant spectral regions to assess how the model can adapt to noise by assessing contextual cues from neighboring regions. In total, four noise levels were assessed at 5\% noise, 10\% noise, 20\% noise, and 30\% noise for both tested methods. Robustness of predictions was assessed using a \SI{0.5}{\micro\molar} MAE cutoff.

For comparison with CRIME, feature importance and model explainability was assessed using Logic Explained Networks (LEN)\cite{Ciravegna2021}, and Shapley Additive Explanations (SHAP)\cite{Lundberg2017}.  The LEN was implemented using PyTorch in python. In order to apply the LEN, the spectral input data was sectioned to discrete categories, and concept mapping was done through taking the mean of the min-max scaled feature map activations across each layer of the model. Each concept corresponded to approximately 25 x-axis points across the SERS spectrum corresponding to approximately half a peak. Further details of the LEN implementation are described in \textbf{Supplementary Material section \ref{sec:LEN}.} SHAP calculations were done using the above-mentioned sectioned categories separately using Gradient Explainer.

\section{Results}
\label{headings}

\subsection{Denoising autoencoder}
In the present work, the utility of deep learning methods was assessed for identifying target serotonin concentrations from SERS measurements with CB[8] additives as chemical spacers. To this end, a denoising autoencoder and neural network quantification models were developed using 682 spectral measurements taken from water and artificial urine-based samples, with concentrations of serotonin ranging from 0 to 9 µM.  Following training, the denoising autoencoder was able to robustly reconstruct the clean data from noisy inputs in the test-set (MSE=0.025). Examples of input noisy data and subsequently denoised spectra are presented in \textbf{Supplementary Figure \ref{supp_fig:1}A}. Similar trends are observed in denoising experimental urine measurement spectra, which have been presented in \textbf{Supplementary Figure \ref{supp_fig:1}B}.

\subsection{Quantification models}
\begin{figure}[htbp]
    \centering
    \includegraphics[width=1\linewidth]{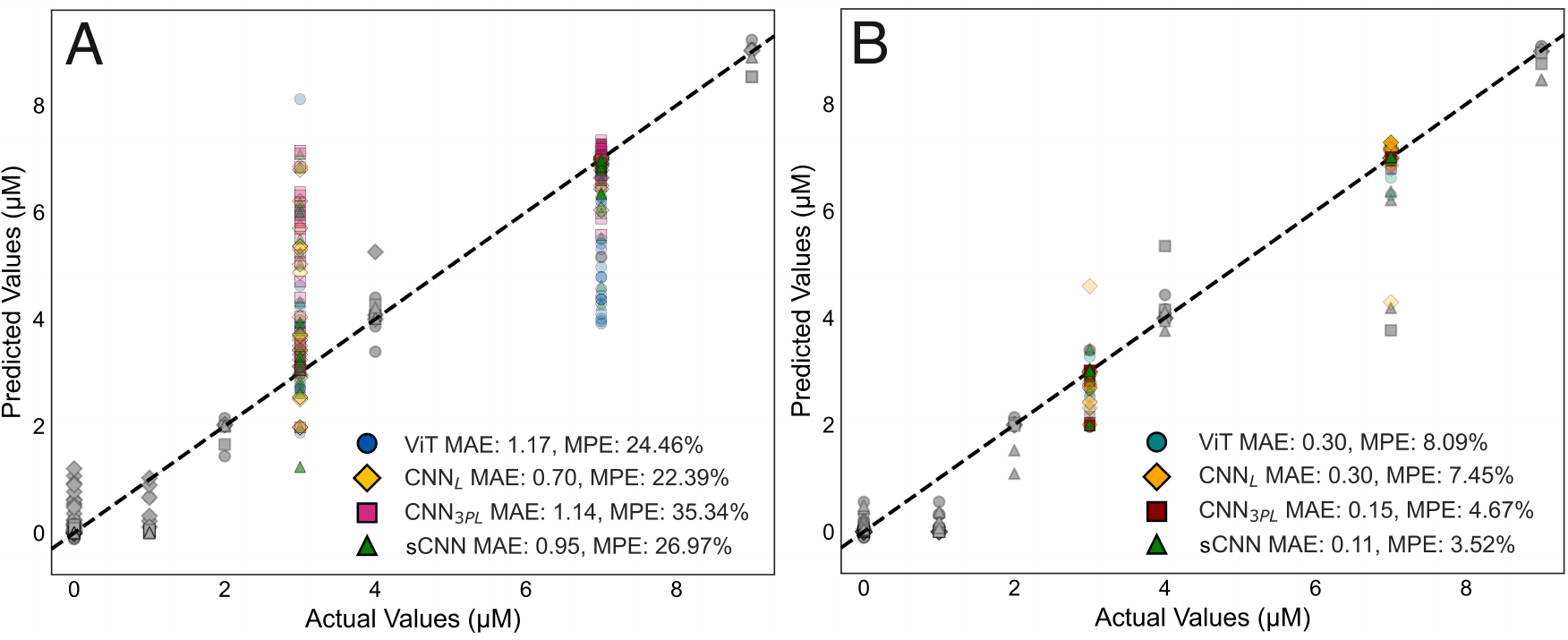}
    \caption{\textbf{Results of the final models in the validation and test sets for the four model types in both raw (A) and denoised datasets (B).} Validation set results are shown in grey, and test set results are shown in color: the linear CNN model is shown in yellow (diamond), the vision transformer model in blue (circle), the scale-adjusting CNN in green (triangle), and the three-parameter logistic output layer CNN model in red (square). The shown values were obtained from the final test set. Validation set results are presented in Table \ref{valresults} in the appendix. MAE = mean absolute error, MPE = mean percentage error.}
    \label{fig:4}
\end{figure}

Four different neural network models were evaluated in both raw and denoised datasets. These included the ViT model, the linear output layer CNN model (CNN\(_{L}\)), the three-parameter logistic output layer CNN model (CNN\(_{3PL}\)), and the scale-adjusting three-parameter logistic output layer CNN model (sCNN). All ensemble models followed a similar trend in predictions in the validation set and are presented in \textbf{Supplementary Figure \ref{supp_fig:2}}. Final selected best model validation set predictions for all three model types are presented in \textbf{Figure \ref{fig:4}} and \textbf{Supplementary Table \ref{valresults}}, and for both datasets, all four models showed strong performance in the validation set. The models were then applied on the test set which comprised of unseen concentration combinations by either the denoising autoencoder or the neural network quantification models. The performance of the selected best models is presented in \textbf{Figure \ref{fig:4}B}. None of the models were able to reach satisfactory differentiation of serotonin from the other neurotransmitters in the raw urine dataset (ViT: MAE~=~\SI{1.17}{\micro\molar}, MPE~=~24.46\%, CNN\(_{L}\): MAE~=~\SI{0.70}{\micro\molar}, MPE~=~22.39\%, sCNN: MAE~=~\SI{0.95}{\micro\molar}, MPE~=~26.97\%, CNN\(_{3PL}\): MAE~=~\SI{1.14}{\micro\molar}, MPE~=~35.34\%). However, in the denoised dataset, all models were capable of robust quantification of serotonin, with the CNN\(_{3PL}\) model (MAE~=~\SI{0.15}{\micro\molar}, MPE~=~4.67\%) and the sCNN model (MAE~=~\SI{0.11}{\micro\molar}, MPE~=~3.52\%) outperforming both the ViT model (MAE~=~\SI{0.30}{\micro\molar}, MPE~=~8.09\%) and the CNN\(_{L}\) model (MAE~=~\SI{0.30}{\micro\molar}, MPE~=~7.45\%). 

\subsection{CRIME framework}

\begin{figure*}[htbp]
    \centering
    \begin{adjustbox}{center}
    \includegraphics[width=1.4\linewidth]{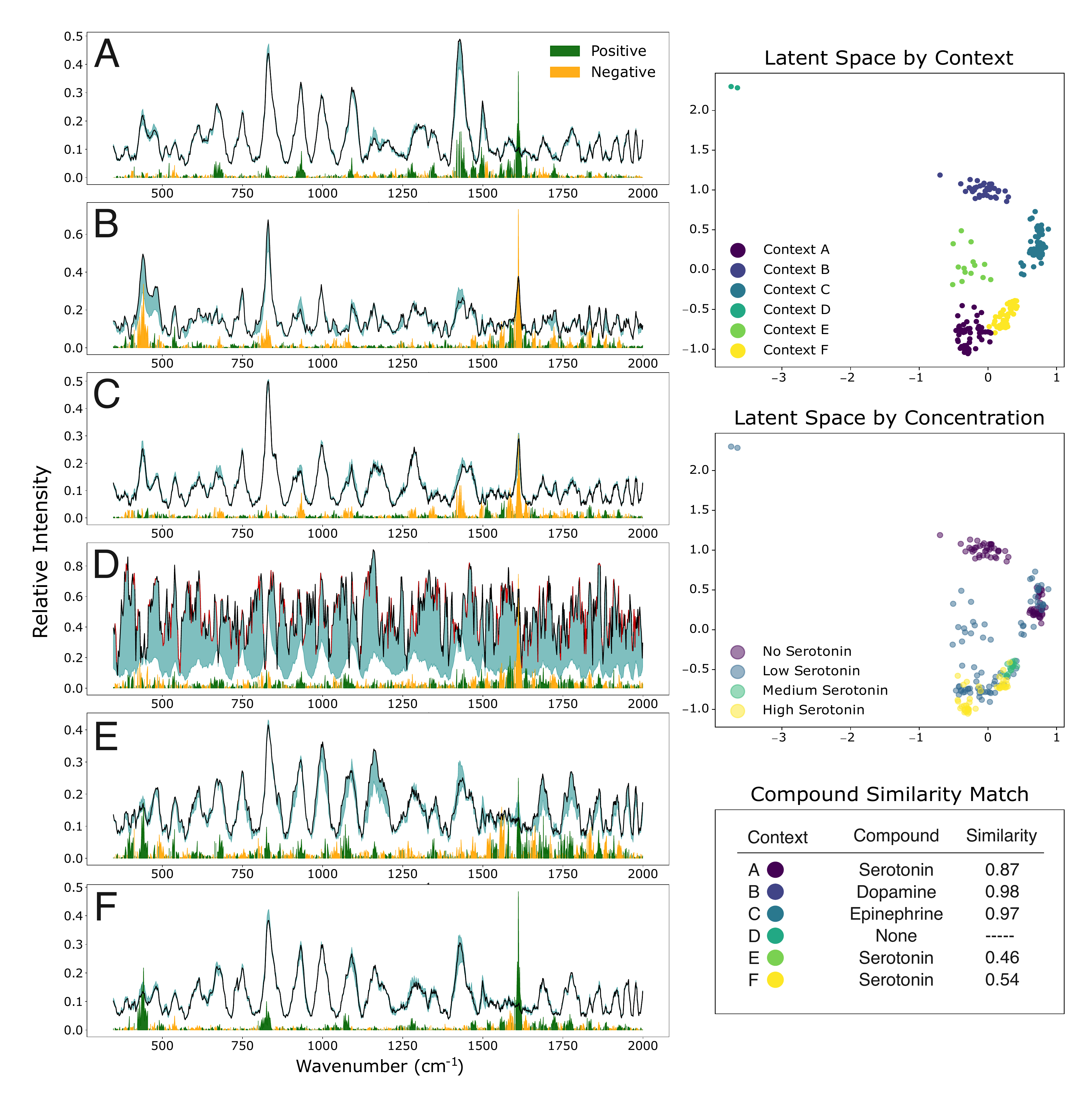}
    \end{adjustbox}
    \caption{\textbf{Results for Context Representative Interpretable Model Explanations (CRIME) analysis.} Six distinct contexts were identified, which are visualized across mean spectra in subfigures A - F. Positive prediction weights are presented in green, negative prediction weights in yellow, and perturbation limits have been shaded in teal. Red regions in the mean spectra correspond to average perturbation limits at either the top or bottom of the feature weight range for the simplicity of the plot. Latent spaces are visualized by context and concentrations, and compound similarity matching was done using cosine similarity. The highest similarity score is presented alongside the matched compound.}
    \label{fig:6}
\end{figure*}

The CRIME framework was fit on a VAE using the LIME explanations of the CNN\(_{3PL}\) predictions. This setting was selected due to its strong performance across both the validation and test data. Following the K-means clustering, the latent space of the VAE was clustered into six contexts. The latent space is visualized in \textbf{Figure \ref{fig:6}}. Among them, four distinct contexts were identified (contexts A, B, C, and F), as well as one intermediate context (context E), and one outlier context (context D). The mean LIME explanations for each CRIME context cluster are presented in \textbf{Figures \ref{fig:6}A} to \textbf{\ref{fig:6}F}. Peak regions were selected further from the most prominent clusters in a three-dimensional representation of the spectral intensities, explanation weights, and position z-scores. The peak-region cluster plots are visualised in \textbf{Supplementary Figures \ref{supp_fig:3}A to \ref{supp_fig:3}F}. Contexts A (S\(_{cos}\)=0.87), E (S\(_{cos}\)=0.46), and F (S\(_{cos}\)=0.54) were correctly associated with serotonin, while contexts B and C were associated with dopamine (S\(_{cos}\)=0.98) and epinephrine (S\(_{cos}\)=0.97) respectively. Complete cosine similarity values between mean CRIME context spectra and reference neurotransmitter spectra are presented for each CRIME context in \textbf{Supplementary Table \ref{table3}}.

\subsection{Benchmarking}

 The neural network models were benchmarked between different architectures and against other machine learning models, with the results summarized in \textbf{Table \ref{mainresults}} and \textbf{Supplementary Figure \ref{XGB}}. The benchmarking results of the CNN architectures are presented in \textbf{Supplementary Table \ref{results}}. Of all the benchmarked models, the PLSR model trained with autoencoder-denoised spectra showed the best performance amongst non-neural network-based models with a MAE of 0.70 \SI{}{\micro\molar}, a sixfold error compared to the sCNN. Models trained with denoised spectra near universally showed superior performance when compared to raw spectra trained models, or second derivative Savitzky-Golay denoised spectra trained models with the exception of the random forests model. Lastly, perturbation testing revealed the final neural network quantification model to be robust to noise, with model predictions crossing the set threshold (MAE < 0.5 \SI{}{\micro\molar}) at 30\% localized noise, and 10\% universal noise.

\begin{table}[ht]
  \centering
  \caption{\textbf{Comparison of test-set mean absolute errors across different machine learning models for serotonin quantification from SERS spectra.} Best performing model MAEs within each dataset has been bolded, and the two best models overall have been marked with an asterisk (*). Additionally, for baseline comparison, the mean absolute error of the previously published PLSR model has been presented. sCNN = convolutional neural network with scaling layers and three parameter logistic (3PL) output layer, CNN\(_{3PL}\) = convolutional neural network with 3PL output layer, CNN\(_{L}\) = convolutional neural network with linear output layer, SVM = support vector machine, RF = random forests, PLSR = partial least squares regression, XGB = extreme gradient boosting, ViT = vision transformer.}
  \label{mainresults}
  \begin{tabular}{lcccccccc}
    \hline
    \bf{Dataset} & \bf{XGB} & \bf{PLSR} & \bf{RF} & \bf{SVM} & \bf{CNN}\(_{3PL}\) & \bf{CNN}\(_{L}\) & \bf{sCNN}& \bf{ViT} \\
    \hline
    Denoising Autoencoder & 0.78 & 0.70 & 0.93 & 0.96 & 0.15* & 0.30 & \bf{0.11*}& 0.30 \\
    Raw Spectra & 0.82 & 2.05 & 0.88 & 1.26 & 1.14 & \bf{0.70} & 0.95& 1.17 \\
    Savitzky-Golay & 1.15 & 2.25 & 1.46 & 1.37 & - & - & - & - \\
    Kasera et al. 2014 & - & \bf{0.52} & - & - & - & - & - & - \\
    \hline
  \end{tabular}
\end{table}

To compare the context explanations to methods of global explainability, LEN and SHAP frameworks were evaluated as a reference standard. The mean feature activations of the CNN\(_{3PL}\) model across all layers is presented in \textbf{Supplementary Figure \ref{supp_fig:4}}. The LEN identified logic statements explaining the four categories of serotonin concentrations with fair (0.69, no serotonin) to excellent (0.98, medium serotonin) explanation accuracy. The logic statements are visualized in \textbf{Supplementary Figures \ref{supp_fig:no}-\ref{supp_fig:high}}. Peak regions near wavenumbers of 800, 1000, 1200, and 1450 cm\textsuperscript{-1} were consistently selected within the first-order-logic (FOL) statements for all concentration ranges and were deemed to be relevant for serotonin concentration prediction. SHAP values were assessed for all concentration ranges separately and have been visualized on an averaged spectra in \textbf{Supplementary Figure \ref{supp_fig:SHAP}}.

\section{Discussion}

Within the present study, a comprehensive framework of spectral quantification from complex biological media was developed, consisting of data preprocessing and denoising, bio-quantification, and model explanation through the CRIME framework. To this end, data from 318 spectra from lyophilized urine media, as well as 364 spectra from water media were utilized for the development of neural network models for denoising of urine backgrounds and quantification of serotonin. The trained denoising autoencoder improved prediction outcomes near-universally across all model types and enabled robust quantification. The assessed state-of-the-art neural network models substantially outperformed traditional machine learning methods commonly used in the SERS domain, with the CNN\(_{3PL}\) and the scale-adjusting sCNN models yielding the lowest prediction errors. Notably, the models developed in the present study substantially outperformed existing methodologies\cite{Kasera2014}. Additionally, the developed CRIME explainability framework identified the spectral contexts in which the model was reliably assessing the relevant serotonin peaks, as well as contexts representing confounding factors or other sample artefacts, which were not readily observable from the outputs of the LEN or the SHAP explanations.

It can be assessed that the custom layers developed in the present study for the sCNN and CNN\(_{3PL}\) models significantly improved quantification performance. The final output layer of a model acts as a type of calibration curve to the final transformations from the input data. In this sense, a regression task of biomolecular quantification is effectively a calibration task. The use of logistic calibration curves can often yield a better fit on data compared to linear curves, as near the limit of quantification an assay can become saturated, or as the values approach the limit of detection, the linearity of the signal can deteriorate. The scaling layers similarly were able to improve the predictions, most likely due to their added capability to handle variation in scale. This was surprisingly relevant despite present samples all being measured using an identical setup, spectrometer, chemicals, and location which would suggest limited reasons for significant changes in intensity scaling. These layers should be further assessed in different SERS tasks to confirm their utility in subsequent studies.

The CRIME framework within this study was able to effectively explain the CNN model exhibiting a significant improvement in the understanding of the model decisions. The found contexts were associated with the neurotransmitters present in the mixture, and it could be assessed that the two largest contexts were accurately representing serotonin, while two contexts were associated with unwanted signals, and one context was ambiguous in its associations. The dopamine- and epinephrine-associated contexts reveal the imbalances present within the dataset, as the misidentified contexts were primarily in samples with low or absent serotonin concentrations. Following this observation, it can be concluded that within the present dataset, a correlation existed between a lack of serotonin and the presence of other neurotransmitters. Therefore, expanding the dataset to include samples with low concentrations of all neurotransmitters could remedy the apparent inability of the model to generalize to lower serotonin concentrations. When compared directly to the explanations provided by the trained LEN and SHAP explainers, it can be seen that there is a robust association of peaks to serotonin in all three methods. However, SHAP explanations could not effectively communicate the potential presence of confounding factors. Similarly, LEN statements, while effective in identifying potential confounders, were complex enough to make their presentation unintuitive. However, this could be remedied through a similar context-seeking variation of LEN. Within applications where the identification of all prediction reasoning is crucial, the application of the CRIME framework could see benefit over current best practices for global explainability. 

It must be highlighted, that the CRIME framework combined with SERS could see clinically relevant use through acting as the first step in biomarker discovery trials. Instead of assessing individual biomarkers of disease through established hypotheses, a biomarker discovery study could be initiated in a non-targeted, hypothesis generating fashion. Applying a machine learning model on raw spectra presently is not advisable due to the lack of confidence in the model assessing true biomarkers as opposed to confounding factors. The exact identification of the signalling biomarkers is challenging when global explainability methods are used for peak detection, as the spectral signals could be a result of multiple overlapping compounds. However, were the CRIME framework applied, individual target biomarkers could be identified through contexts uniquely, and subsequently assigned to the likely biomarkers through a complete library of present compounds, as well as hypothesized biomarkers. With the advent of computational hypothesis generating methodologies such as Mendelian randomization\cite{Zaki2023}, future biomarker discovery trials could see a significant change in early-stage methodology and approach. Similarly, effects of potential comorbidities or medication on spectral signals could be identified through association of contexts to such groupings, and therefore their effects could be mitigated within a biomarker analysis. For example, in a diagnostic task predicting schizophrenia, there would be a significant risk of misidentifying the effects of comorbidities and antipsychotic medications as disease biomarkers. However, following a CRIME analysis, different contexts could be identified corresponding to these effects or their known biomarkers. Various strategies could then be applied to remove these effects from the spectra. To validate the CRIME framework's explainability and clinical utility, ongoing efforts are being directed towards a biomarker study, which aims to establish the framework's effectiveness in relevant clinical scenarios.

There are several limitations to consider in this study. The development of a denoising autoencoder in patient urine samples as opposed to artificial urine samples could prove to be more challenging. While it could be explained by the CRIME framework, the neural network models were not always able to assess the association of the peaks to the target serotonin compound directly, and instead assessed the presence of dopamine or epinephrine as a predictor of the absence of serotonin.  The CRIME framework in turn presents limitations inherent in LIME explanations, as the explanations are dependent on a simpler model fit to complex model predictions and as such do not completely represent an explanation of the actual model. Additionally, it could prove challenging to identify the true reasoning behind CRIME contexts if there were more potential confounding compounds or effects present. Lastly, while in the present task it was feasible to calculate the LIME explanations for all instances, this might not be applicable in larger datasets.

In conclusion, the present study set out to develop a machine learning predictive approach which was capable of achieving three main aims: to perform enhanced single analyte detection of serotonin from a mixture of varying neurotransmitter concentrations in urine, to recognize and adjust for method-based variation in SERS measurements, and lastly to identify all prediction contexts applied by the quantification model. A denoising autoencoder was developed to improve the targeting of relevant neurotransmitter peaks. Additionally, to assess serotonin quantification in raw and denoised spectra, three different state-of-the-art neural network models were developed: a CNN with a three parameter logistic output layer, a scale-adjusting CNN, and a ViT model. In addition, all models were compared to other machine learning methods. Finally, a novel model explainability framework, CRIME, was built around LIME explanations through the assessment of prediction contexts using a combination of VAE and clustering algorithms. The model interpretability was compared between the novel framework and global prediction methods: LEN, and SHAP. Within this study, it was shown that the denoising autoencoder substantially improved predictive capabilities of applied machine learning models, of which the developed three-parameter logistic output layer CNN outperformed other models assessed. Moreover, model explainability was strongly enhanced through the CRIME framework. To our knowledge, this marks the first instance where an autoencoder has been successfully applied to biological ‘noise’ within the SERS domain. Within the chemical spectral domain, the CRIME framework promises to enable deployment of spectral quantification methods to directly identify disease features in biological fluids, which could be further refined into specific biomarkers through the identification of relevant contexts.

\subsection*{Acknowledgements}
The authors would like to thank Dr Setu Kasera for their detailed and elaborate data collection enabling the repurposing of previously measured SERS spectra for the present study.

\subsection*{Availability of Data and Materials}
The developed code for the CRIME framework can be found in the following GitHub repository: \href{https://github.com/jkz22/CRIME}{https://github.com/jkz22/CRIME}

\subsection*{Funding}
This work was supported by the Stanley Medical Research Institute (grant number: O7R-1888) by grants to Sabine Bahn, and by the Oskar Huttunen Foundation grant to Jihan K. Zaki.

\subsection*{Conflicts of Interest}
The authors have no conflicts to declare.

\subsection*{Author Contribution Statement}
Conceptualization: JKZ, PL, OAS; \\
Methodology: JKZ, PL;   \\
Data analysis: JKZ; \\
Resources: SB, OAS; \\
Writing - Original Draft: JKZ;  \\
Writing - Review \& Editing: All co-authors;    \\
Supervision: PL, SB, OAS, JT;                   \\
Funding acquisition: OAS, JKZ               \\

\bibliographystyle{plain}
\bibliography{library}

\newpage

\appendix

\setcounter{figure}{0}
\setcounter{table}{0}

\section*{Supplementary Material}

\section{Experimental Methods}
\label{sec:Experimental}

All initial reagents were sourced from Alfa Aesar and Merck and were utilized in their received state unless otherwise specified. Cucurbit[8]uril was prepared following established literature protocols. Millipore water with a resistivity of 18 M $\Omega$·cm was employed in all experiments, unless otherwise indicated. Fresh standard stock solutions of neurotransmitters, specifically dopamine, epinephrine, and serotonin, were prepared at varying concentrations to simulate potential interfering background analytes. Gold nanoparticles (AuNP) with a diameter of 60 nm stabilized by citrate were procured from British Biocell International (BBI). Lyophilized urine samples designated as Calibrator Lot No. 150 and Control Level II Lot No. 230 were obtained from RECIPE ClinChek-Control and were reconstituted in dilute hydrochloric acid as per the supplier's guidelines. 

Spectra for both Raman and SERS were collected with a 785 nm laser operating at 17.5 mW, using an Ocean Optics QE65000 Spectrometer. Each spectrum was acquired over a 10 s interval. AuNPs with a 60 nm diameter were first centrifuged at 12,000 rpm for 45 s, repeated twice, and \SI{900}{\micro\liter} of the supernatant were removed. Subsequently, a sample preparation sequence was followed: neurotransmitters (dopamine, epinephrine, and serotonin) were first added, followed by \SI{50}{\micro\liter} of the centrifuged AuNPs, then \SI{20}{\micro\liter} of CB[8] at a final concentration of \SI{20}{\micro\molar}, and finally, \SI{50}{\micro\liter} of thawed urine, which had been initially stored on ice. An identical procedure was replicated, replacing urine with water for control experiments.

\section{Neural network architectures}

\label{sec:NNQM}
Both the CNNs and the ViT models were implemented in TensorFlow and designed to adapt to SERS spectral data. The CNN architecture comprised sequential layers optimized for 1D convolution operations, and the core CNN architecture was used in all trained CNN models, where the initial layer is a convolutional layer featuring 8 filters and a large kernel size approximately the width of half a peak (25 datapoints), aimed to capture broader features in the spectrum. This initial layer employs a Rectified Linear Unit (ReLU) activation function and reduces sequence length through striding. Intermediate layers employ paired hyperbolic tangent (Tanh) and ReLU activation functions, designed to capture complex patterns while maintaining non-linearity. The combination of consequential Tanh and ReLu layers is to direct the model to assess the upper half of the identified general peaks from the sweeper layer. These layers maintain the same padding to avoid changes in sequence length. The model contains two Tanh-ReLu paired layers with 2~×~16 and 2 ~×~32 filters respectively, with a filter size of 9. The final convolutional stage employs 64 filters with a smaller kernel size of 2 using ReLU activation, aimed to capture fine-grained details in the data. Subsequently, the data is flattened and passed through two fully connected layers employing the same Tanh to ReLu structure with 32 and 16 nodes respectively, to serve the regression task. The core architecture of the CNN models was benchmarked against similar architectures with Tanh-ReLu pairs replaced with ReLu pairs, inversed ReLu-Tanh pairs, or single ReLu layers. Each of the benchmark architectures were trained and validated as described in the methods section, and the results are summarized in \textbf{Supplementary Table \ref{results}}.

\begin{table}[ht]
  \centering
  \caption{\textbf{Comparison of core CNN architectures.} All architectures were trained on denoised urine medium datasets, and validated using the holdout test set. ReLu = Rectified linear unit, Tanh = Hyperbolic tangent, MAE = Mean absolute error, MPE = Mean percentage error.}
  \label{results}
  \begin{tabular}{lcccc}
    \hline
    \bf{Error} & \bf{Tanh-ReLu} & \bf{ReLu-Tanh} & \bf{2xReLu} & \bf{ReLu} \\
    \hline
    MAE & 0.30 & 0.56 & 0.88 & 0.78\\
    MPE & 7.45 & 17.68 & 20.71 & 16.61\\

    \hline
  \end{tabular}
\end{table}

\begin{table}[H]
  \centering
  \caption{\textbf{Validation set results for neural network models.} sCNN = convolutional neural network with scaling layers and three parameter logistic (3PL) output layer, CNN\(_{3PL}\) = convolutional neural network with 3PL output layer, CNN\(_{L}\) = convolutional neural network with linear output layer, ViT = vision transformer.}
  \label{valresults}
  \begin{tabular}{lcccc}
    \hline
    \bf{Dataset} & \bf{CNN}\(_{3PL}\) & \bf{CNN}\(_{L}\) & \bf{sCNN}& \bf{ViT} \\
    \hline
    Denoising Autoencoder & 0.24 & 0.13 & 0.33 & 0.20 \\
    Raw Spectra & 0.19 & 0.31 & 0.14 & 0.25 \\
    \hline
  \end{tabular}
\end{table}

\section{AI explainability}
\label{sec:AIx}

\subsection{CRIME variational autoencoder}
\label{sec:CRIME}

Within the encoder of the CRIME VAE, the input data \(X\), is transformed into the mean (\(\mu\)) and logarithm of the variance (log(\(\sigma^2\))) in a proposed Gaussian distribution in the latent space through a fully connected ReLu layer with 256 nodes. During training, the outputs of the encoder are passed to a sampling layer which generates a random noise variable \(\epsilon\) generated from a standard gaussian distribution, which is then transformed using the encoder outputs to draw samples \(z\) as such: \(z = \mu + \sigma \ast \epsilon\). The decoder then applies a mirrored dense layer network to the encoder with a ReLu layer with 256 nodes, and a final output sigmoid layer with 3~×~842 nodes. The model was trained for 128 epochs and a batch size of 32, with the Adam optimizer with a learning rate of 0.001 and using the sum of the mean squared error, and Kullback–Leibler divergence as the loss function. 

\begin{table}[htbp]
  \centering
  \caption{\textbf{Cosine similarity values across explanation weighted reference spectra and explanation weighted mean context spectra.} Highest similarity values within a context cluster are bolded. X = context. }
  \label{table3}
  \begin{tabular}{lcccccc}
    \hline
    \bf{Reference compound} & \bf{A}& \bf{B}& \bf{C}& \bf{D}& \bf{E}& \bf{F}\\
    \hline
    Serotonin   & \textbf{0.87}& 0.85      & 0.60       & -0.79        & \textbf{0.46} & \textbf{0.54} \\
    Dopamine    & 0.05& \textbf{0.98}& 0.91       & -0.80   & 0.29      & 0.06      \\
    Epinephrine & 0.0& 0.81& \textbf{0.97}  & -0.79        & 0.12      & 0.08      \\
    \hline
  \end{tabular}
\end{table}

\subsection{Logic explained network architecture}
\label{sec:LEN}

The LEN architecture was modified to be similar to the original model architecture, consisting of an entropy layer with 164 input nodes, a leaky ReLu layer with 32 nodes, a Tanh layer with 16 nodes, a ReLu layer with 4 nodes, and a final linear output layer. The LEN was trained using weight decay Adam as the optimizer, using binary cross entropy with logits loss as the loss function with a scaled auxiliary entropy loss at a 0.000001 multiplier. The model was trained with 5001 epochs using a learning rate of 0.0001. 

\section{Benchmark hyperparameter search}
\label{sec:Benchmarks}

The grids used for the hyperparameter search of the benchmark machine learning models are presented below with the final hyperparameters for the denoised models in bold. 

\subsection*{XGBoost}
\begin{tabular}{ll}
\toprule
Hyperparameter       & Values                             \\
\midrule
colsample\_bytree    & \textbf{0.5}, 0.7, 0.8                  \\
learning\_rate       & \textbf{0.01}, 0.1, 0.2, 0.3               \\
max\_depth           & 3, \textbf{6}, 9, 12                 \\
alpha                & 1, \textbf{3}, 5                           \\
n\_estimators        & 100, 300, 600, 900, \textbf{1200} \\
\bottomrule
\end{tabular}

\subsection*{Random Forests}
\begin{tabular}{ll}
\toprule
Hyperparameter       & Values                          \\
\midrule
max\_depth           & 1, 2, 3, 6, 7, \textbf{8}, 10           \\
n\_estimators        & 100, 300, 600, 900, \textbf{1200} \\
\bottomrule
\end{tabular}

\subsection*{PLSR}
\begin{tabular}{ll}
\toprule
Hyperparameter       & Values \\
\midrule
n\_components        & 5, 8, \textbf{12} \\
\bottomrule
\end{tabular}

\subsection*{SVM}
\begin{tabular}{ll}
\toprule
Hyperparameter       & Values                          \\
\midrule
C                    & 0.1, 1, 10, \textbf{50}, 100            \\
epsilon              & \textbf{0.01}, 0.1, 1                   \\
gamma                & \textbf{scale}, auto                    \\
\bottomrule
\end{tabular}

\section*{Supplementary Figures}
\label{sec:suppFig}

\begin{figure}[ht]
    \centering
    \includegraphics[width=0.9\linewidth]{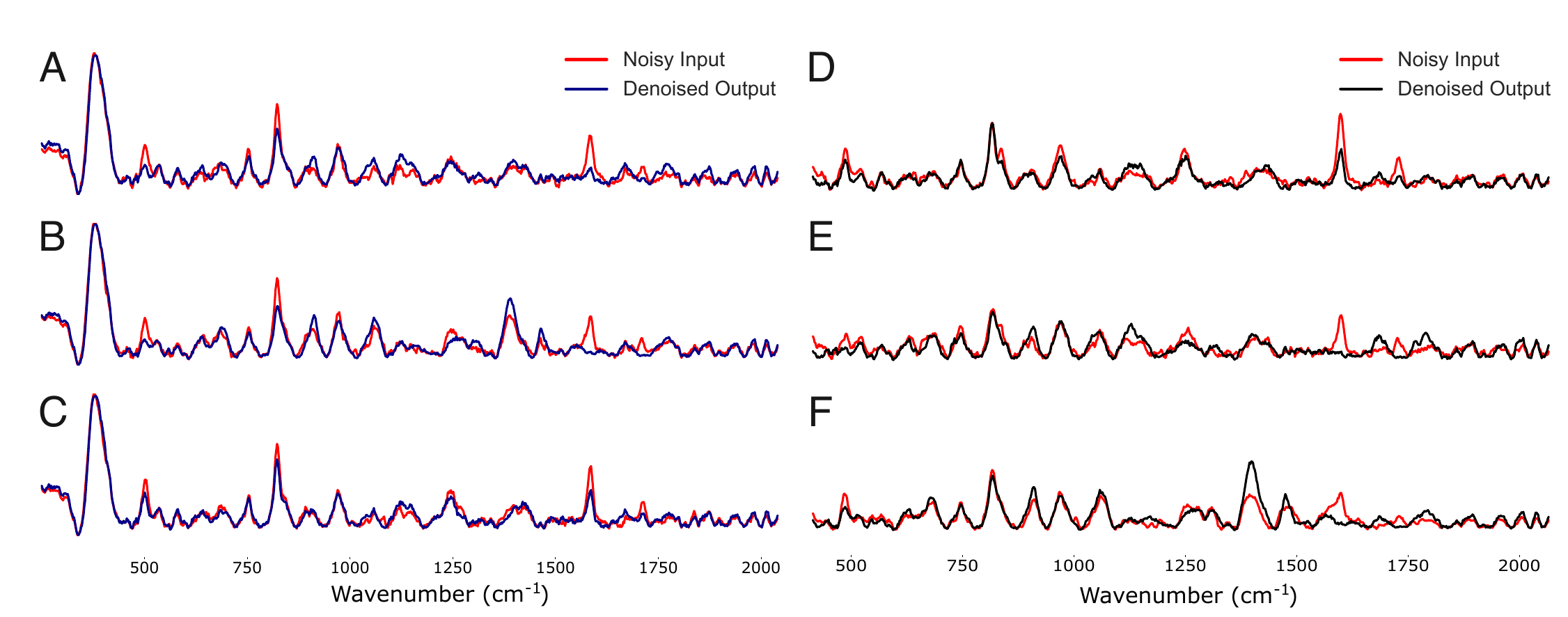}
    \caption{\textbf{Examples of denoised spectra in the artificial training data (A-C), and in lyophilized urine spectra (D-F).}Y-axis (relative intensity) is omitted for clarity.}
    \label{supp_fig:1}
\end{figure}

\begin{figure}[ht]
    \centering
    \includegraphics[width=1\linewidth]{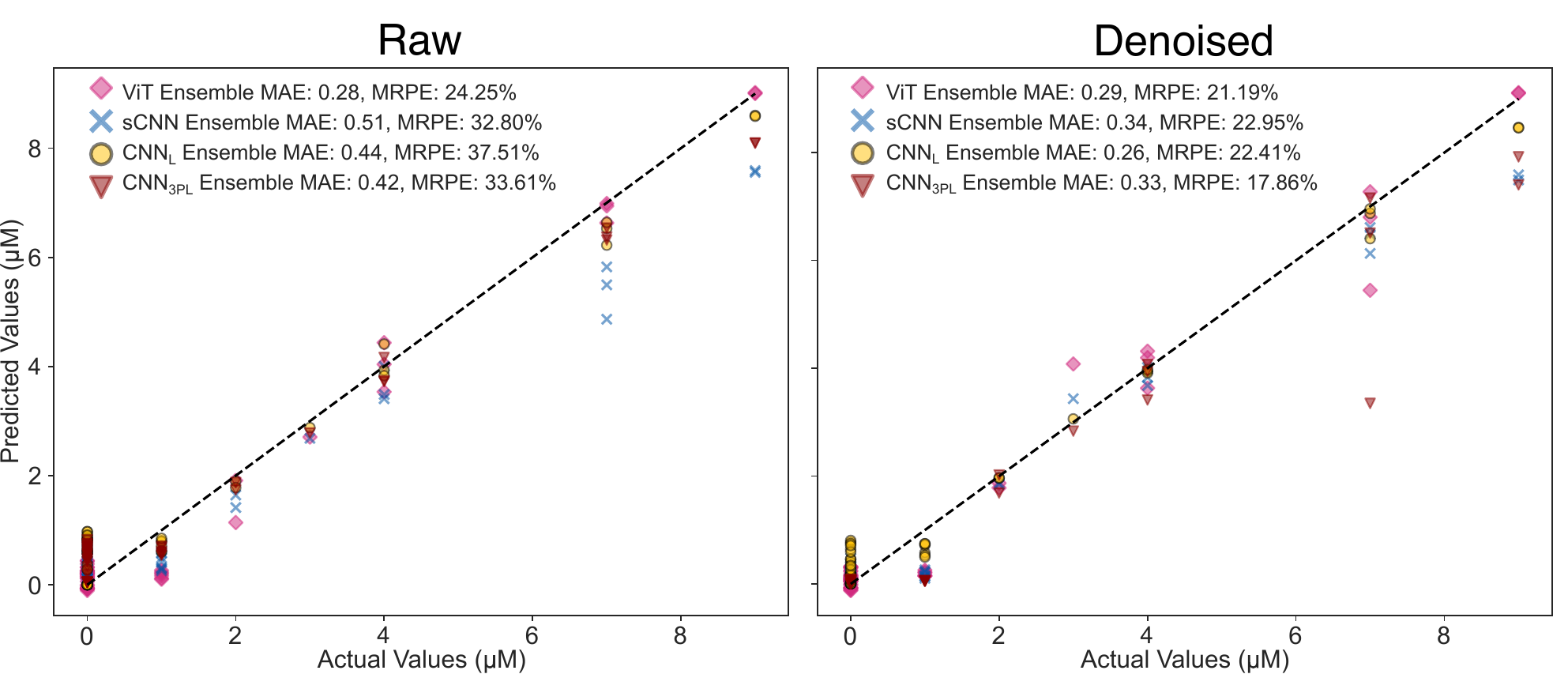}
    \caption{\textbf{Predictions of the trained ensembles on the validation set for all types of neural networks on both raw spectra and denoised spectra.}}
    \label{supp_fig:2}
\end{figure}

\begin{figure}[ht]
    \centering
    \includegraphics[width=0.8\linewidth]{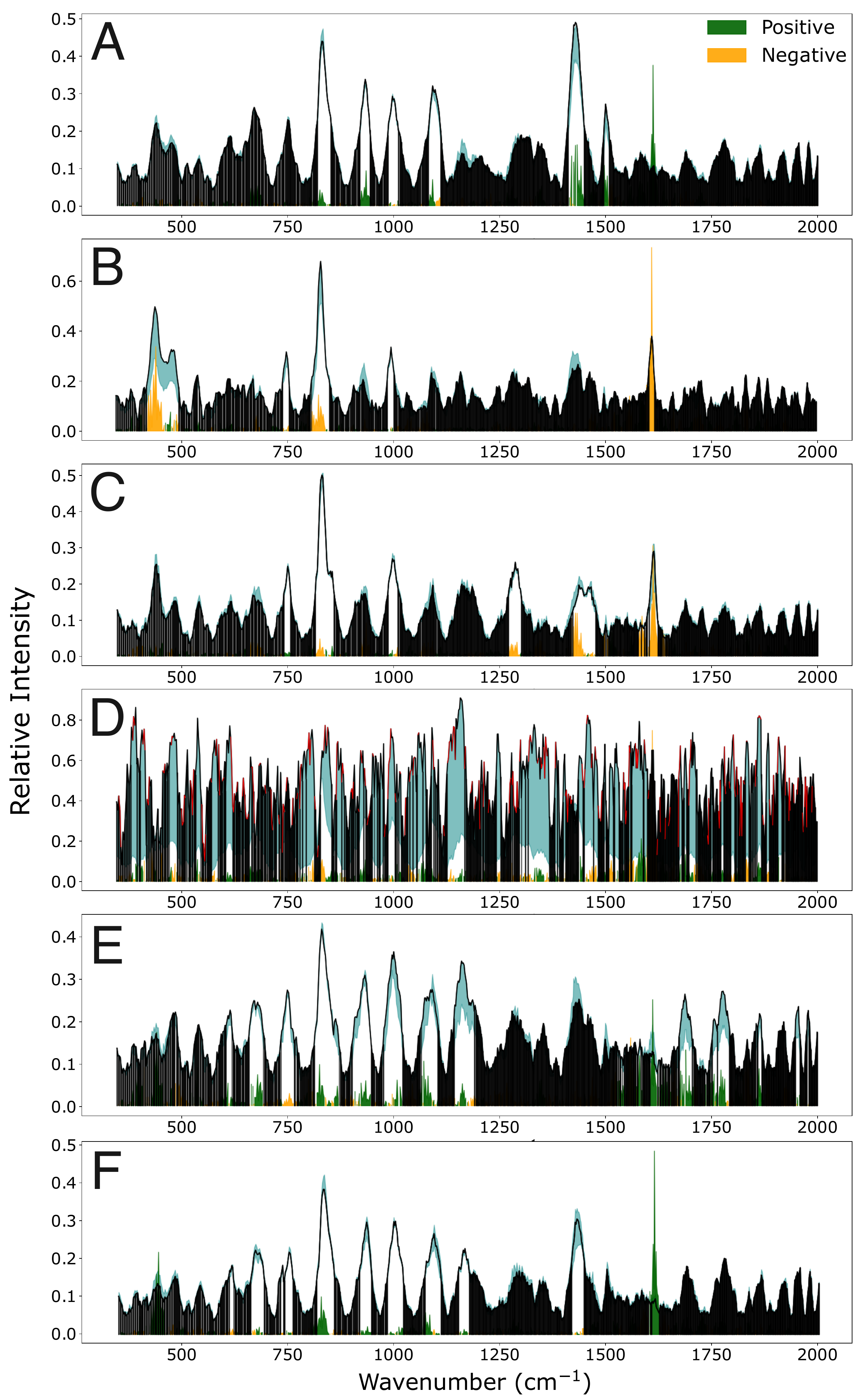}
    \caption{\textbf{Peak-region clusters of high relevance extracted from CRIME contexts for compound identification.} Context labels correspond to labels in Figure \ref{fig:6}. Positive prediction weights are presented in green, negative prediction weights in yellow, and perturbation limits have been shaded in teal. Red regions in the mean spectra correspond to average perturbation limits at either the top or bottom of the feature weight range for the simplicity of the plot. Areas not relevant are marked in black. High-relevance clusters were obtained with K-means clustering of the product of peak height and LIME weights, with the top 5 largest clusters selected.}
    \label{supp_fig:3}
\end{figure}

\begin{figure}[ht]
    \centering
    \includegraphics[width=1\linewidth]{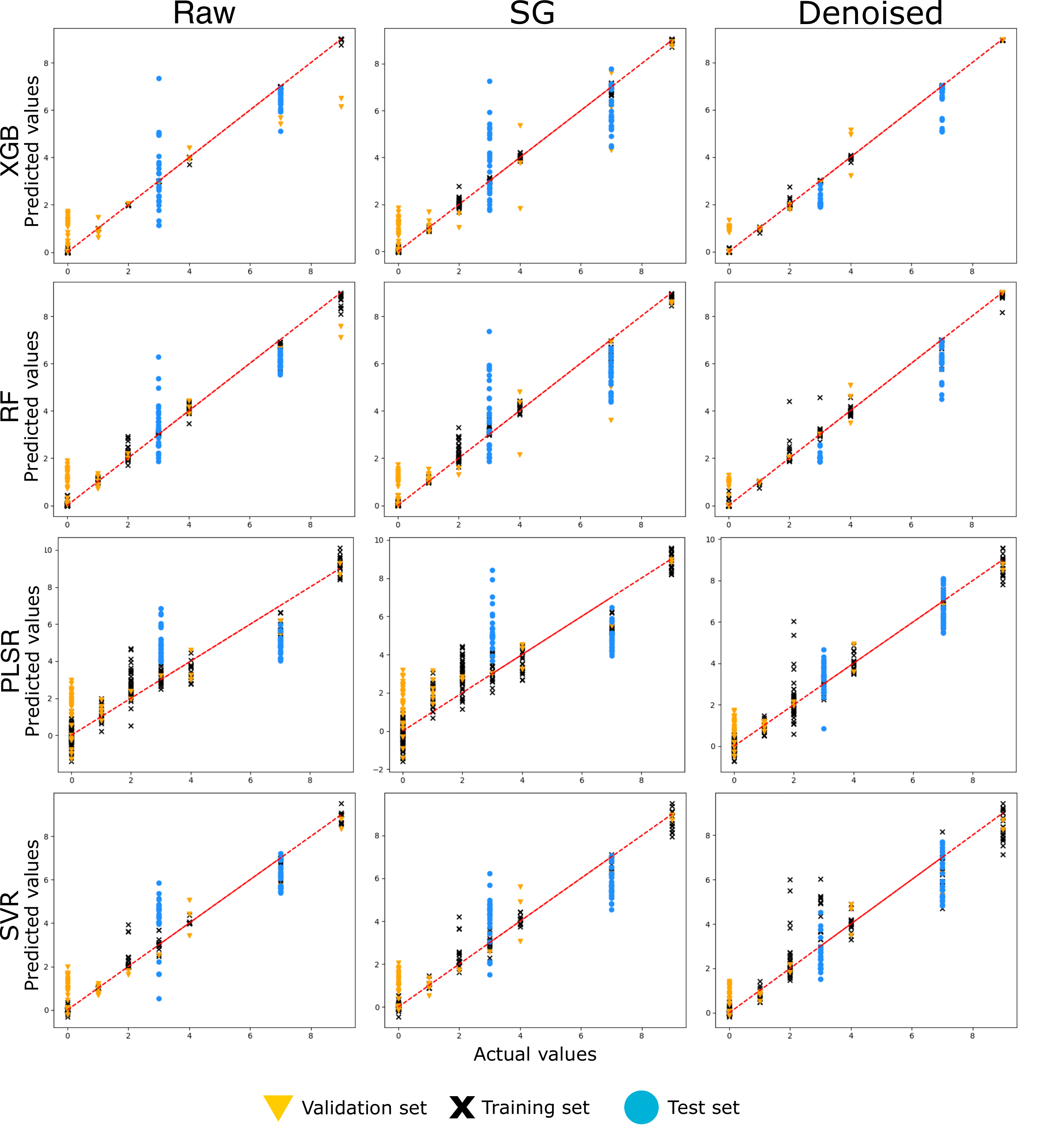}
    \caption{\textbf{Results for quantification model benchmarking.} Training set predictions are marked in black (cross), validation set predictions in orange (triangle), and test set predictions in blue (circle). XGB = extreme gradient boosting, RF = random forests, PLSR = partial least squares regression, SVM = support vector machine regression, SG = Savitzky-Golay filter.}
    \label{XGB}
\end{figure}

\begin{figure}[ht]
    \centering
    \begin{adjustbox}{center}
    \includegraphics[width=1.4\linewidth]{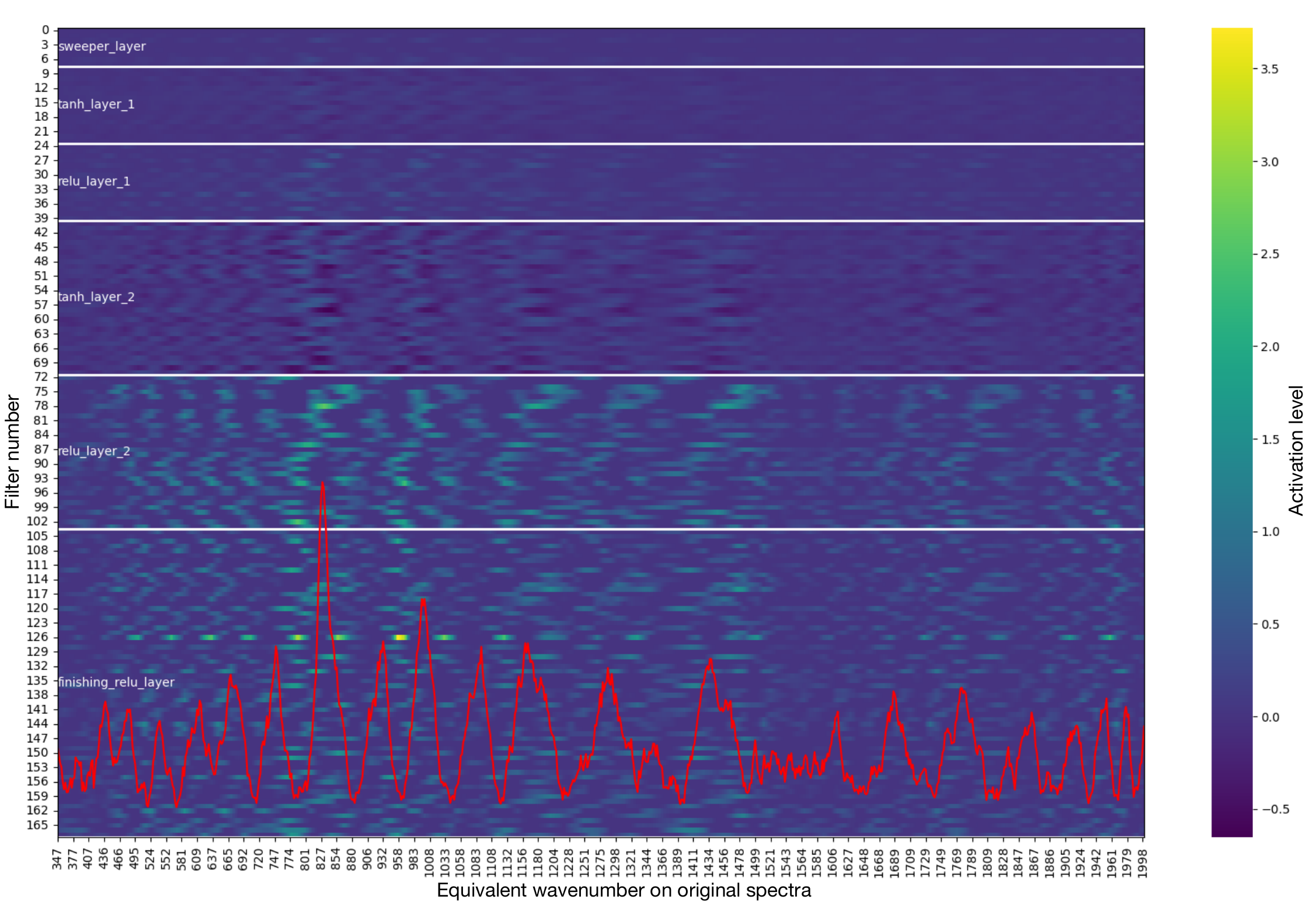}
    \end{adjustbox}
    \caption{\textbf{Feature activation map for each convolutional layer in the CNN model overlayed with an example spectra.} SERS spectra is shown in red, and higher activations are marked with a yellow hue, with lower activations marked in blue. }
    \label{supp_fig:4}
\end{figure}

\begin{figure}[ht]
    \centering
    \includegraphics[width=0.8\linewidth]{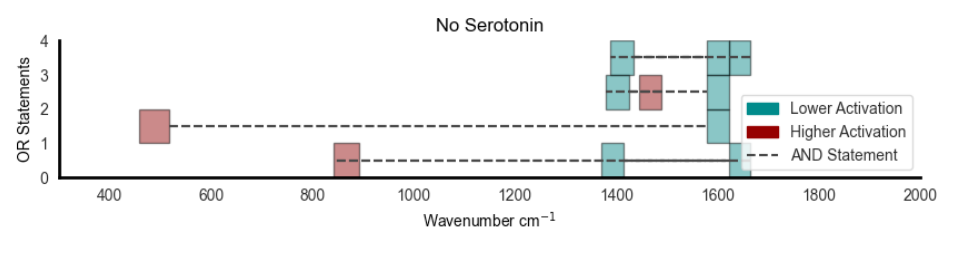}
    \caption{\textbf{LEN results visualized for samples with no serotonin concentration.} OR statements are separated vertically, and AND statements are presented level with a dashed line connecting the statements. Blue squares denote lower activation and red squares higher activation.}
    \label{supp_fig:no}
\end{figure}

\begin{figure}[ht]
    \centering
    \includegraphics[width=0.8\linewidth]{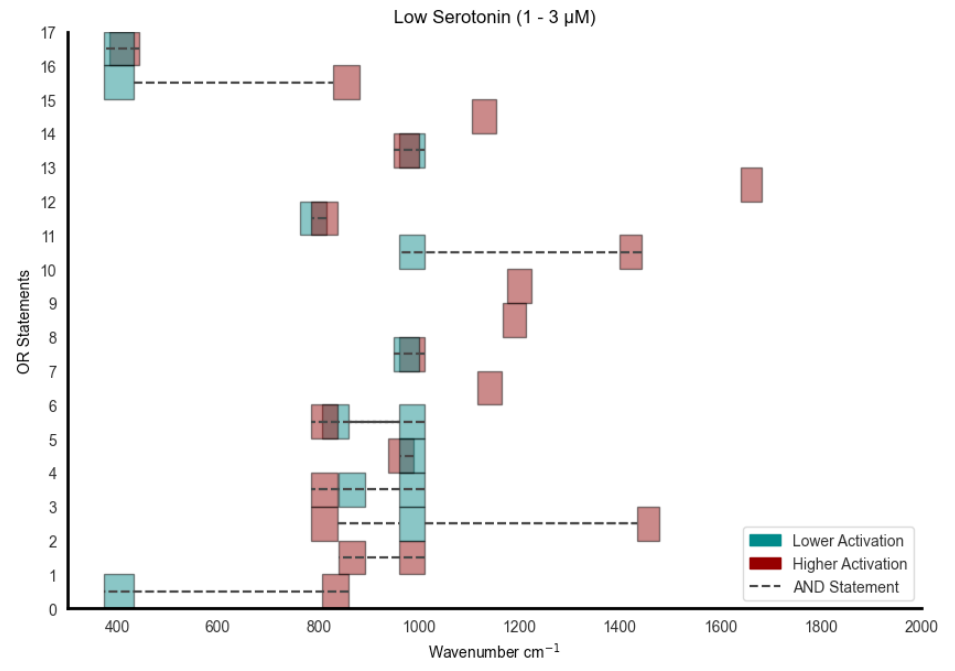}
    \caption{\textbf{LEN results visualized for low serotonin concentrations.} OR statements are separated vertically, and AND statements are presented level with a dashed line connecting the statements. Blue squares denote lower activation and red squares higher activation.}
    \label{supp_fig:low}
\end{figure}

\begin{figure}[ht]
    \centering
    \includegraphics[width=0.8\linewidth]{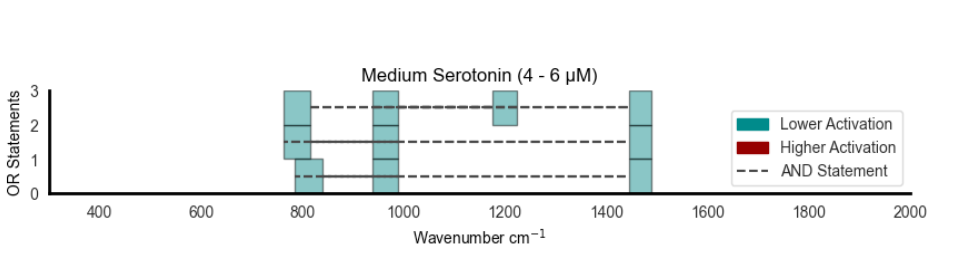}
    \caption{\textbf{LEN results visualized for medium serotonin concentrations.} OR statements are separated vertically, and AND statements are presented level with a dashed line connecting the statements. Blue squares denote lower activation and red squares higher activation.}
    \label{supp_fig:medium}
\end{figure}

\begin{figure}[ht]
    \centering
    \includegraphics[width=0.8\linewidth]{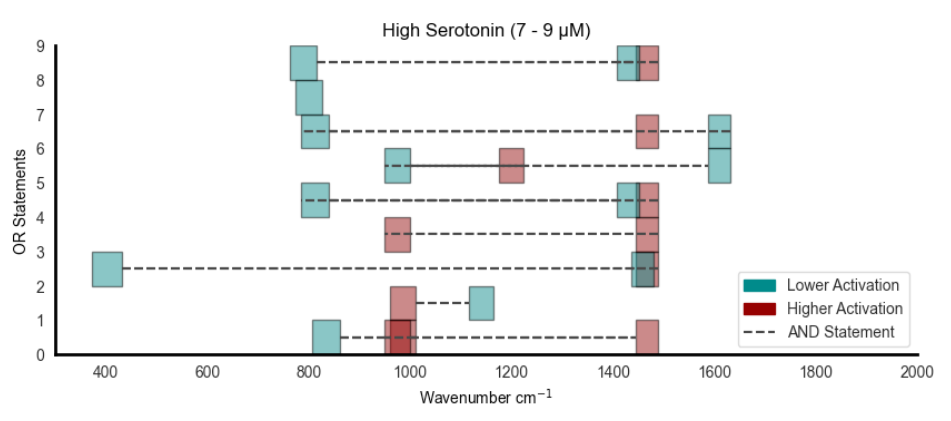}
    \caption{\textbf{LEN results visualized for high serotonin concentrations.} OR statements are separated vertically, and AND statements are presented level with a dashed line connecting the statements. Blue squares denote lower activation and red squares higher activation.}
    \label{supp_fig:high}
\end{figure}

\begin{figure}[ht]
    \centering
    \includegraphics[width=1\linewidth]{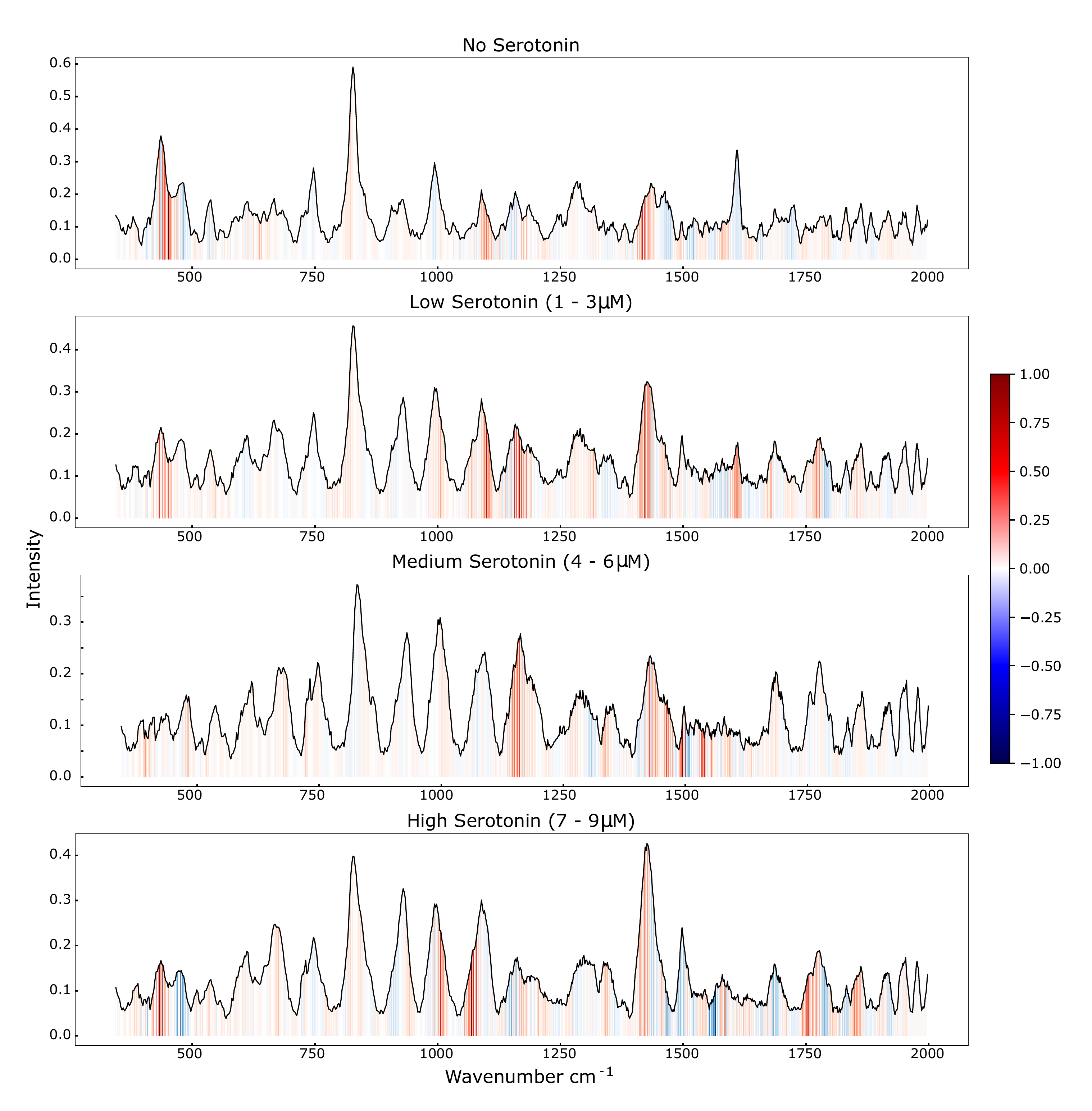}
    \caption{\textbf{Shapley additive explanations (SHAP) visualized for all serotonin concentration ranges.} Spectra shown are mean spectra across the respective concentration ranges. SHAP values were obtained using Gradient Explainer, and red areas correspond to positive SHAP values and blue areas to negative SHAP values.}
    \label{supp_fig:SHAP}
\end{figure}

\end{document}